# Topologically-protected refraction of robust kink states in valley photonic crystals

Fei Gao[1#], Haoran Xue[1#], Zhaoju Yang[1*], Kueifu Lai[2,3], Yang Yu[2], Xiao Lin[1], Yidong Chong[1,4,*], Gennady Shvets[2,*], Baile Zhang[1,4,*]

[1]Division of Physics and Applied Physics, School of Physical and Mathematical Sciences, Nanyang Technological University, Singapore 637371, Singapore.

[2]School of Applied and Engineering Physics, Cornell University, Ithaca NY 14853.

[3]Department of Physics, University of Texas at Austin, Austin, TX 78712, USA.

[4]Centre for Disruptive Photonic Technologies, Nanyang Technological University, Singapore 637371, Singapore.

#Authors contribute equally to this work.

*Author to whom correspondence should be addressed; E-mail: zhaojuyang@ntu.edu.sg (Z. Yang); yidong@ntu.edu.sg (Y. Chong); gshvets@cornell.edu (G. Shvets); blzhang@ntu.edu.sg (B. Zhang)

**Recently discovered[1,2] valley photonic crystals (VPCs) mimic many of the unusual properties of two-dimensional gapped valleytronic materials[3-11] such as bilayer graphene[5,7-10] or $MoS_2$[11]. Of the utmost interest to optical communications is their ability to support topologically protected chiral edge (kink) states[3-5] at the internal domain wall between two VPCs with spectrally overlapping bandgap zones and opposite half-integer valley-Chern indices. We experimentally demonstrate the robustness of the kink states in VPCs that support degenerate transverse-electric-like (TE) and transverse-magnetic-like (TM) topological phases, thus enabling polarization multiplexing in a single topological waveguide. The propagation direction of the kink states is locked to the valleys of the reverse Brave lattice and, therefore, cannot be reversed in the absence of inter-valley scattering. At the intersection between the internal domain wall and the external edge separating the VPCs from free space, the kink states are shown to exhibit >97% out-coupling efficiency into directional free-space beams. This constitutes the first experimental demonstration of meron-like valley-projected topological phases with half-integer valley-Chern indices.**

The valley is a new binary degree of freedom (DOF) occurring in lattices with certain crystal symmetries; it is used in 'valleytronics'[3-11] as a novel way to transport information and energy, similar to how spin is used in spintronics[12]. Compared to spintronic topological edge states, which are protected by time-reversal symmetry in quantum spin Hall (QSH) systems, topological valley transport is protected by crystal symmetries that forbid valley mixing[9], and thus is experimentally challenging[9,10] to realize in condensed matter systems because inter-valley scattering is ubiquitous. On the other hand, the valleytronic approach does not require a strong spin-orbit interaction[13], and can be actively controlled in various 2D materials (e.g., bilayer graphene) by a spatially dependent electric field[5,7,10] that determines the sign of the Berry curvature at the valleys of the high-symmetry (K and K') points of the Brillouin zone. This enables the creation of distinct valley phases separated by internal domain walls, or 'kinks', which are populated by one-dimensional topologically protected 'kink states'[3-5].

In the emerging field of topological photonics[14-25], the appeal of using the valley DOF is even greater because of the challenges in realizing a synthetic spin DOF. All proposed and experimentally realized approaches to developing photonic analogues of a spin, such as using directionally coupled rings[16-18], or creating a superposition of two degenerate polarization states[19-24], have considerable disadvantages such as size or complexity. Recent theories that introduced the valley DOF into photonic crystals[1,2] suggest that valley topological transport can provide a simpler route to achieving robust propagation of confined photons.

In this Letter, we experimentally demonstrate a VPC (see Fig.1) that acts as a photonic analogue of the quantum valley Hall topological insulator[9,10], supports robust kink states, and possesses the following unique features. First, by utilizing the decoupled TE/TM polarization DOF, two pairs of topological valley kink states can be selectively excited at the domain wall between two VPCs. This doubles the number of topologically protected transmission channels, and is an advantage over the recently-reported topological valley transport of sound[26], which was limited by the longitudinal nature of air-borne compressional waves. Second, the topologically-protected kink states can out-couple, or refract, with near-perfect efficiency into free space, through specific valley-preserving terminations of the VPC. Such topologically protected refraction is a phenomenon that is specific to electromagnetic waves. It represents a major distinction of VPCs from their electronic valley-Hall topological insulator counterparts because the latter do not allow electrons to refract into vacuum. It is also fundamentally different from the common light refraction at Brewster angle, which applies only to plane waves with a specific polarization. Third, we experimentally show that valley conservation is the precondition for topologically protected transport and refraction of the kink states. This is accomplished by deliberately constructing an armchair termination that breaks valley conservation and causes backscattering of kink states. Such a controlled experiment is virtually impossible in condensed matter systems.

As depicted in Fig. 1a, the designed VPC is a triangular lattice whose unit cell consists of a metallic tripod suspended between two parallel metallic plates[24] (see Supplementary Information for detailed design procedure). The simulated bulk band diagram is shown in Fig. 1b. Because the tripod geometry breaks the inversion symmetry, a band gap (5.8 GHz $< f <$ 6.2 GHz) emerges near the K(K') valleys for both the TE and TM bands, which are degenerate at the two valleys. The eigenmode profiles around the K valley (for the bands labelled '1' to '4' in Fig. 1b) are plotted in Fig. 1c. For the TE eigenmodes, the Poynting vector rotates clockwise/counter-clockwise around the tripod for the eigenmode of the bands '1' and '4'. This shows that the photonic valley DOF corresponds to an orbital angular momentum, similar to the valley DOF in electronic systems. For the TM eigenmodes, the rotation of the Poynting vector switches to be counter-clockwise/clockwise in the empty region among the tripods for the eigenmode of band '2'/'3'. This shows that the polarization can indeed act as an independent DOF for the original valley pseudospin. Moreover, the band topology analysis shows the valley-Chern indices[3,27,28] are half-integer: $C_K = 1/2$ and $C_{K'} = -1/2$ for both TE/TM polarizations (see Methods). This implies that, while the existence of the edge states at the interface between a VPC and free space is not guaranteed[29], the kink states emerge between the VPCs with the opposite signs of the valley-Chern index.

Now we construct a 'kink'-type domain wall between two VPCs with opposite valley-Chern indices. As shown in Fig. 2a, the lower domain (the same as in Fig. 1) has valley-Chern index $C_{K(K')}=\pm 1/2$. The upper domain has tripods oriented in the opposite direction as in Fig. 1, and thus exhibits opposite valley-Chern index $C_{K(K')}=\mp 1/2$. The difference in valley-Chern indices across the domain wall indicates that there will be two topological kink states whose propagation directions are locked to the K and K' valleys (band diagram simulated in Fig. 2b) owing to the half-integer topological indices[27,28] of the valley-projected topological phases of a VPC that have a meron-type fractional topological charge[30-32]. Note that this is a major distinction of the VPC from that of the electronic valley-Hall topological insulators, such as electrically gated bilayer graphene, which actually possesses four kink states (two per valley) at the valley-preserving zigzag interface because they possess integer valley Chern numbers[5,8].

Vertical (horizontal) single dipole antennas oriented along the $z$ ($y$) direction were placed at the left end of the domain wall to launch the TM (TE) polarized waves. Figures 2c and 2d show the transmitted $H_z$ (TE) and $E_z$ (TM) fields for the two polarizations. Both transmission bands are consistent with the simulated kink state bands in Fig. 2b. The field distributions scanned along the $y$ direction (Figs. 2e and 2f) show that the kink states for both polarizations are indeed confined at the domain wall.

As the band gap is tunable by adjusting the angle $\alpha$ (Fig. 1a), we rotate the tripods in the upper domain till $\alpha = 60°$ in order to close the band gap (see Supplementary Information), thus destroying the kink states. Consequently, the two transmission bands in Figs. 2c and 2d drop dramatically.

To demonstrate robust valley transport of the kink states in absence of inter-valley scattering inside the VPC, we constructed a zigzag path for the domain wall, as shown in Fig. 3a. For both polarizations, the measured transmission (Figs. 3b and 3c) shows negligible suppression in the band gap (5.8-6.2 GHz), and is comparable to the transmission along a straight domain wall of equal length. We plot the simulated $H_z$ field pattern of TE polarization and $E_z$ field pattern of TM polarization at 6.03 GHz in Figs. 3d and 3e. The kink states can be guided around the zigzag path without reflection.

In order to experimentally demonstrate topologically-protected refraction of the kink states at the valley-preserving zigzag termination of the VPC into free space (see the geometry in Figs. 4e and 4f), this is accomplished by launching a single-valley (thus, unidirectional) kink states propagating towards the free space termination. We refer to the space on the other side of the interface as 'free space', since it is free of photonic structure, but note that it is confined between the top and bottom plates. Because there is only one kink state per valley, two phased dipoles positioned along the domain wall is sufficient for generating single-valley states (see Supplementary Information for the setup and measurement details). The reflectance results for the TE (Fig.4a) and TM (Fig.4c) modes are shown as the light red curves. Nearly-vanishing ($R_z < 3\%$) reflectance is observed across the entire band gap when the termination is zigzag-shaped, i.e. preserves the valley DOF.

The experimentally scanned free space field patterns at $f = 6.12\text{GHz}$ plotted in Fig.4e ($H_z$ for the TE mode) and Fig.4f ($E_z$ for the TM mode) reveal that the refracted beams are highly directional, and in excellent agreement with the simulation results plotted in the same figures. By comparing the two wave polarizations, we observe one important distinction: the TE mode refracts in a single direction while the TM mode refracts into two nearly-orthogonal directions. In order to interpret this surprising behaviour, we have developed a simple semi-analytic model that predicts the refraction directions of the kink states existing from the VPC. The model is based on applying phase matching conditions at the terminal interface, as shown in Fig. 4b (for the TE mode) and Fig.4d (for the TM mode), between the free-space modes and the kink states that are assumed to be weakly localized. The right-moving kinks state for both polarizations are locked to the K' valley, as marked by three black dots $\vec{K}'_i$ (where $i = 1,2,3$) at the equivalent corners of Brillouin zone.

On the other hand, the free space dispersion relations are different for the two polarizations: $k_{\text{TE}} = \sqrt{(\omega/c)^2 - (\pi/d)^2}$ and $k_{\text{TM}} = \omega/c$, as illustrated by the red/green circles inside/outside the Brillouin zones shown in Figs.4b,d. Therefore, the refractive behaviour of the two polarizations is expected to be different due to the difference in their respective free-space refractive indices. Applying the phase matching condition to the interface parallel to $\vec{e}_{zig}$ requires finding, for each polarization, the free-space wave vectors $\vec{k}$ that satisfy the following conditions:

$\vec{k} \cdot \vec{e}_{zig} = \vec{K}'_i \cdot \vec{e}_{zig}$ and $|\vec{k}| = k_{TE,TM}$. These are graphically solved in Figs. 4b,d. Because $k_{TM} > k_{TE}$, the above equations have two solutions for the TM mode but only one for the TE mode, in excellent agreement with experimental observations and numerical simulations.

As the ultimate demonstration of the necessity of valley conservation for topological protection of the kink states, we deliberately construct an armchair termination of the VPC (Figs. 4g and 4h) in order to break the valley conservation. Thus nonzero reflection is expected. As shown in Figs. 4a and 4c, the measured reflectance is now larger by one order of magnitude than that from the zigzag termination. Note that the relatively large reflectance of TE polarization is because of the insufficient sensitivity of the magnetic probe in measuring small signals.

Thus, we have experimentally demonstrated that the near-perfect refraction of both TE/TM kink states into free space is contingent on the conservation of the valley DOF, which is ensured by the choice of the termination: the zigzag termination preserves the valley DOF while the armchair one does not. Moreover, the TE/TM degeneracy of topological modes can double data capacity by polarization multiplexing to support robust and high-speed wireless and optical data networks[33,34]. Due to the high efficiency (over 97%) of the coupling between the topological modes and free space modes, we anticipate new and intriguing practical applications for directional antennas, lasers, and other communication devices across the electromagnetic spectrum.

# Methods

**Fabrication and Simulation.** The aluminum tripods are fabricated with the wire Electric Discharge Machining (wire EDM) method. The band diagrams are simulated with first-principle electromagnetic simulation softwares COMSOL Multiphysics, where the aluminum tripods used in experiments are modelled as perfectly electric conductor (PEC). The dispersion shown in Fig. 2b was performed with a supercell that contains 10 tripods on each side of the interface. The field patterns are simulated with CST Microwave Studio. For the TM mode excitation, a vertical dipole with length 34mm is placed in the middle of the two parallel plate waveguide. For the TE mode excitation, a horizontal dipole with length 34 mm is placed in the middle of the two parallel plate waveguide.

**Dirac Hamiltonian and valley Chern number.**

Photonic lattices with $C_{6v}$ symmetries are known to possess an extra discrete degree of freedom: the valley, which refers to the proximity of propagating electromagnetic waves to one of the two high-symmetry corners at $K = (4\pi/3a_0, 0)$ and $K' = (-4\pi/3a_0, 0)$ of the Brillouin zone. Under a broad set of perturbations[1] that do not scatter photons from one valley into another, the valley is conserved. Under the valley conservation assumption, it becomes appropriate to consider a restricted topological phase of photons that is defined in only one of the two valleys. Such phases are commonly referred to as merons[30-32], and are characterized by a restricted (valley projected) half-integer Chern numbers associated with their valley: $C_K = 1/2$ and $C_{K'} = -1/2$ obtained by integrating the Berry curvature over a restricted region of the Brillouin zone that coincides with one of the valleys. The bulk-boundary correspondence principle prohibits edge states at the interface between a meron topological phase and a topologically trivial phase. However, the kink states at the domain wall between VPCs with half-integer spin-valley Chern numbers of the opposite sign are allowed.

Formally, the band topology of the VPC can be described by a massive Dirac Hamiltonian $H = v_D(\delta k_x \tau_z s_0 \sigma_x + \delta k_y \tau_0 s_0 \sigma_y) + m\tau_0 s_0 \sigma_z$. Here, $v_D$ is the group velocity, ($\delta k_x$, $\delta k_y$) is the momentum deviation from K(K') point, $\sigma_{x,y,z}$, $\tau_z$, are the Pauli matrices acting on orbital and valley state vectors respectively, and $\tau_0$, $s_0$ are unit matrixes acting on valley and polarization state vectors respectively. $m$ is the effective mass induced by inversion-symmetry-breaking of the tripod geometry. This Hamiltonian produces a nontrivial Berry curvature $\Omega$ in the lower TE/TM bands, whose integration near the K(K') valley gives rise to the valley Chern number $C_K=1/2$ and $C_{K'}=-1/2$ for both TE/TM polarizations. Note that the integration of Berry curvature over the whole Brillouin zone is zero because of time-reversal symmetry.

During writing the manuscript, we were aware of two analogous works[35,36].

**Acknowledgements**

This work was sponsored by Nanyang Technological University for NAP Start-up Grants, Singapore Ministry of Education under Grants No. MOE2015-T2-1-070, MOE2015-T2-2-008 and Grant No. MOE2011-T3-1-005. K. L., Y. Y., and G. S. acknowledge the support of the Air Force Office of Scientific Research and of the Army Research Office.

**Author Contributions**

All authors contributed extensively to this work. F. G., H. X. and Z. Y. fabricated structures and performed measurements. F. G., Z. Y., Y. Y. and X. L. performed simulation. F. G. and Z. Y. provided major theoretical analysis. K. L. designed part of the unidirectional excitation experiment. Y. C., G. S., and B. Z. supervised the project.

**Competing Financial Interests statement**

The authors declare no competing financial interests.

# References

1. Ma, T., and Shvets, G., All-Si valley-Hall photonic topological insulator. *New J. Phys.* **18**, 025012 (2016).

2. Dong, J., *et al.* Valley Photonic crystals for control of spin and topology. *Nat. Mater.* **16**, 298 (2017).

3. Yao, W., Yang, S. & Niu, Q. Edge States in Graphene: From Gapped Flat-Band to Gapless Chiral Modes. *Phys. Rev. Lett.* **102**, 096801 (2009).

4. Jung, J., Zhang, F., Qiao Z., & MacDonald, A. H. Valley-Hall kink and edge states in multilayer graphene. *Phys. Rev. B* **84**, 075418 (2011).

5. Qiao, Z., Jung, J., Niu, Q., & MacDonald A. H. Electronic Highways in Bilayer Graphene. *Nano Lett.* **11**, 3453 (2011).

6. Semenoff, G.W., Semenoff, V. & Zhou, F. Domain walls in gapped graphene. *Phys. Rev. Lett.* **101**, 087204 (2008)

7. Martin, I., Blanter, Y. M. & Morpurgo, A. F. Topological confinement in bilayer graphene. Phys. Rev. Lett. 100, 036804 (2008).

8. Zhang, F., MacDonald, A. H. & Mele, E. J. Valley Chern numbers and boundary modes in gapped bilayer graphene. *Proc. Natl Acad. Sci.* USA **110**, 10546-10551 (2013).

9. Ju, L. *et al.* Topological valley transport at bilayer graphene domain walls. *Nature* **520**, 650-655 (2015).

10. Li, J. *et al.* Gate-controlled topological conducting channels in bilayer graphene. *Nat. Nanotech.* **11**, 1060-1065 (2016).

11. Mak, K. F., McGill, K. L., Park, J. & McEuen, P. L. The valley Hall effect in MoS2 transistors. *Science* **344**, 1489-1492 (2014).

12. Zutic, I., Fabian, J., & Sarma, S. Spintronics: Fundamentals and applications. *Rev. Mod. Phys.* **76**, 323 (2004).

13. Kane, C. L., & Mele, E. J. Quantum Spin Hall Effect in Graphene. *Phys. Rev. Lett.* **95**, 226801 (2005).

14. Wang, Z., Chong, Y., Joannopoulos, J. D. & Soljacic, M. Observation of unidirectional backscattering-immune topological electromagnetic states. *Nature* **461**, 772-775 (2009).

15. Rechtsman, M. C. *et al.* Photonic floquet topological insulators. *Nature* **496**, 196-200 (2013).

16. Hafezi, M., Demler, E. A., Lukin, M. D. & Taylor, J. M. Robust optical delay lines with topological protection. *Nature Phys.* **7**, 907-912 (2011).

17. Liang, G. Q. and Chong, Y. D. Optical Resonator Analog of a Two-Dimensional Topological Insulator. *Phys. Rev. Lett.* **110**, 203094 (2013).

18. Gao, F. *et al.* Probing topological protection using a designer surface plasmon structure. *Nat. Commun.* **7**,11619 (2016).

19. Khanikaev, A. B. *et al.* Photonic topological insulators. *Nature Mater*. **12**, 233-239 (2013).

20. Chen, W., Jiang, S., Chen, X., Zhu, B., Zhou, L., Dong, J., and Chan, C. T. Experimental realization of photonic topological insulator in a uniaxial metacrystal waveguide. *Nat. Commun.* **5**, 5782 (2014).

21. Ma, T., Khanikaev, A. B., Mousavi, S. H. & Shvets, G. Guiding electromagnetic waves around sharp corners: topologically protected photonic transport in metawaveguides. *Phys. Rev. Lett.* **114**, 127401 (2015).

22. Xiao, B. *et al.*, Exciting reflectionless unidirectional edge modes in a reciprocal photonic topological insulator medium. *Phys. Rev. B* **94**, 195427 (2016).

23. Lai, K., *et al.* Experimental Realization of a Reflections-Free Compact Delay Lines Based on a Photonic Topological Insulator. *Sci. Rep.* **6**, 28453; doi: 10.1038/srep28453 (2016).

24. Ma, T., and Shvets, G., Scattering-free optical edge states between heterogeneous photonic topological insulators. *Phys. Rev. B* **95**, 165102 (2017).

25. Lu, L. Joannopoulos, J. D., and Soljacic, M. Topological states in photonic systems. *Nat. Phys.* **12**, 626-629 (2016).

26. Lu, J. *et al.* Observation of topological valley transport of sound in sonic crystals. *Nat. Phys.* **13**, 369-374 (2017).

27. Ezawa, M., Topological Kirchholff law and bulk-edge correspondance for valley Chern and spin-valley Chern numbers. *Phys. Rev. B* **88**, 161406 (2013).

28. Ezawa, M., Symmetry protected topological charge in symmetry broken phase: Spin-Chern, spin-valley-Chern and mirror-Chern number. *Phys. Lett. A* **378**, 1180-1184 (2014).

29. Li, J., Morpurgo, A. F., Buttiker, M. & Martin, I. Marginality of bulk-edge correspondence for single-valley Hamiltonians. *Phys. Rev. B* **82**, 245404 (2010).

30. Bernevig, B. A., Hughes, T. L. & Zhang, S.-C., Quantum spin Hall effect and

topological phase transition in HgTe quantum wells. *Science* **314**, 1757-1761 (2006).

31. Qiao, Z., Jiang, H., Li, X., Yao, Y. & Niu, Q., Microscopic theory of quantum anomalous Hall effect in graphene. *Phys. Rev. B* **85**, 115439 (2012).

32. Ezawa, M., Topological phase transition and electrically tunable diamagnetism in silicene. *Eur. Phys. J. B* **85**, 363 (2012).

33. Wellbrock, G., and Xia, T. J., The road to 100G deployment. *IEEE Commun. Mag.* **48**, S14 - S18 (2010)

34. Chen, Z.-Y. *et al.* Use of polarization freedom beyond polarization-division multiplexing to support high-speed and spectral-efficient data transmission. *Light: Sci. Appl.* **6**, e16207 (2017).

35. Noh, J., *et al.* Observation of Photonic Topological Valley-Hall Edge States., arXive preprint arXive: 1706.00059 (2017).

36. Wu, X., *et al.* Direct observation of Valley-polarized Topological Edge States in a Designer Surface Plasmon Crystals., arXive preprint arXive: 1703.04570 (2017).

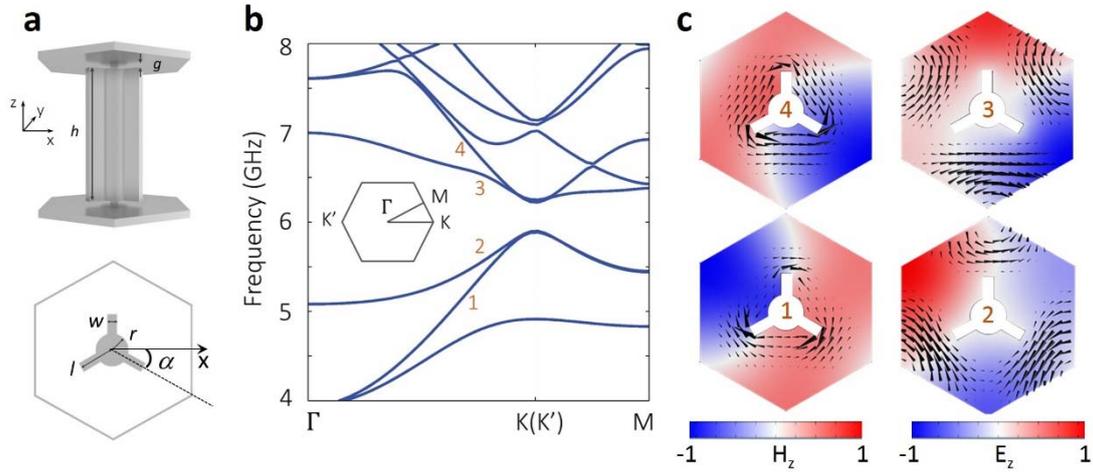

**Figure 1| Topological valley photonic crystal and its bulk band structure.** (a) A unit cell of the lattice, consisting of a metallic tripod suspended in a parallel-plate waveguide. The upper and lower panels are side and top views, respectively. The tripod has height $h$ = 34.6 mm, inner radius $r$ = 3.68 mm, arm length $l$ = 7.95 mm, and arm width $w$ = 2.21 mm. Two air gaps $g$ = 1.1 mm separate the tripod from the upper and lower plates [in experiment the air gaps are filled with a foam spacer (thickness 1.1 mm, ROHACELL@ 71 HF)]. The lattice constant is $d$ = 36.8 mm. (b) The bulk band structure with $\alpha$ = 30°. The inset shows the first Brillouin zone. Bands labelled "1" and "4" have TE polarization, while "2" and "3" have TM polarization. (c) The simulated field patterns of eigen modes in the middle $xy$ plane for corresponding bands "1" to "4", respectively. The black arrows represent Poynting power flows.

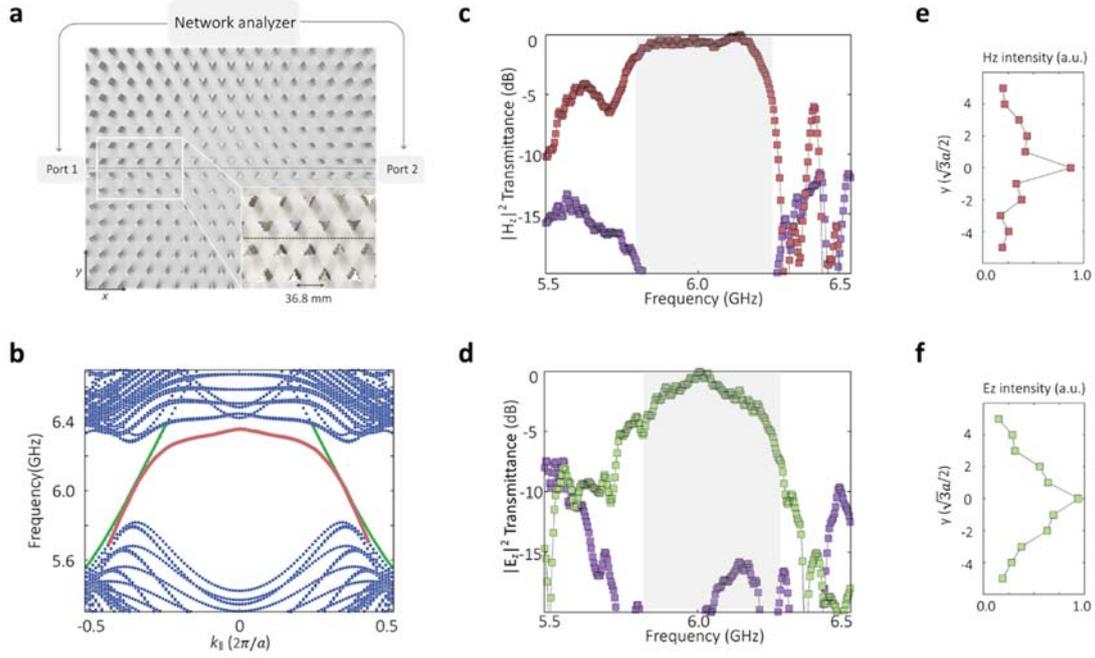

**Figure 2| Selective excitation of topological valley kink states.** (a) The experimental setup for measuring kink states. The straight domain wall is indicated with a dashed line, formed by upper domain with $\alpha = 90°$ and lower domain with $\alpha = 30°$. The inset is a zoomed-in photo. The upper metallic plate of the parallel-plate waveguide is removed for illustration. (b) Band structure of the lattice in a strip geometry. The red and green curves indicate TE and TM polarizations, respectively. (c) Measured $|H_z|^2$ transmittance for the structure in (a) (red), and for a structure of upper domain with $\alpha = 60°$ (purple). (d) Measured $|E_z|^2$ transmittance for the structure in (a) (green) and for a structure of upper domain with $\alpha = 60°$ (purple). (e-f) Decay of kink state along y direction for TE and TM polarizations, respectively.

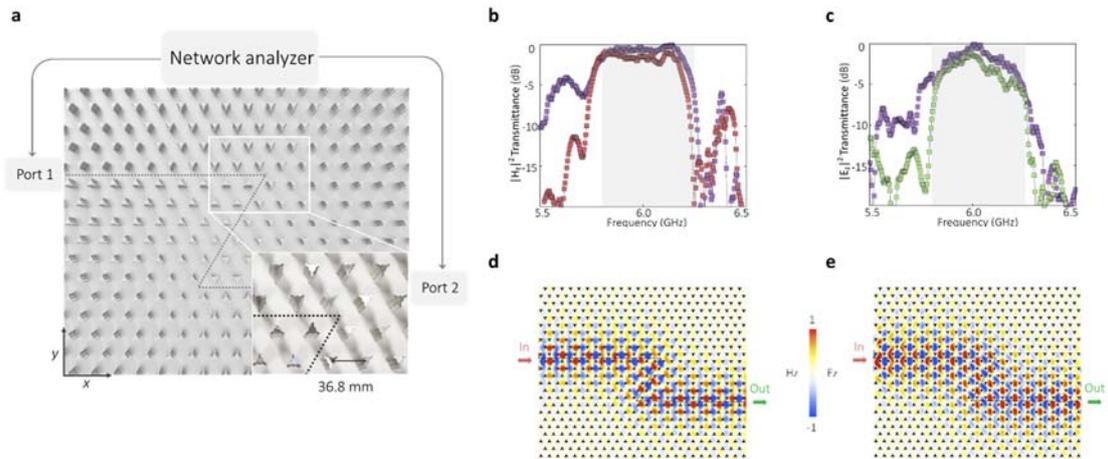

**Figure 3| Robustness of topological valley kink states.** (a) The experimental setup. The black dashed line indicates the zigzag domain wall. The inset is a zoomed-in photo. (b) Comparison of transmittance along a straight (purple) and a zigzag (red) domain walls for TE polarization. (c) Comparison of transmittance along a straight (purple) and a zigzag (green) domain walls for TM polarization. (d) The simulated $H_z$ field pattern for TE polarization at 6.03 GHz. (e) The simulated $E_z$ field pattern for TM polarization at 6.03 GHz.

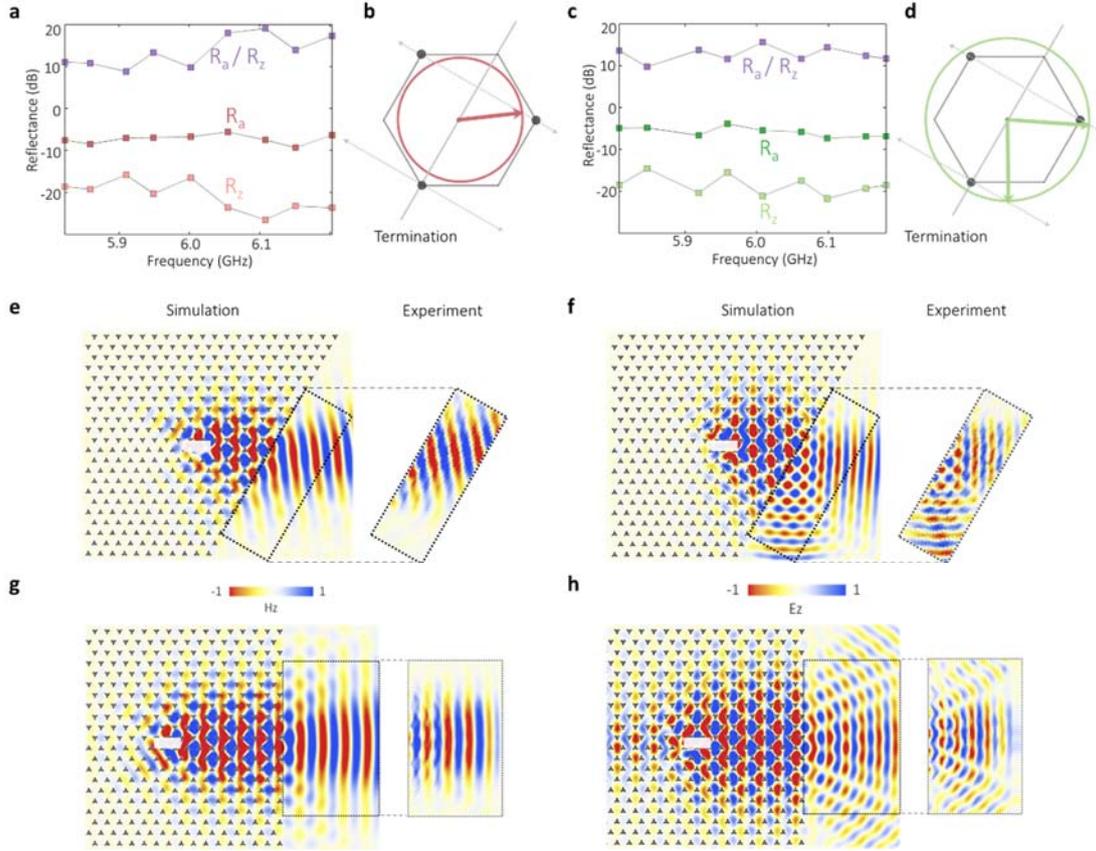

**Figure 4| Topologically-protected refraction of kink states into ambient space.** (a) Measurement of $|H_z|^2$ reflectance for zigzag (light red; labelled "$R_z$") and armchair (dark red; labelled "$R_a$") terminations. Purple line shows the reflectance ratio (labelled "$R_a/R_z$") between the two terminations. (b) The *k*-space analysis on the out-coupling of TE polarization. The red circle represents the TE dispersion in the parallel plate waveguide. The three black dots represent the K' valley in the Brillouin zone. (c) Measurement of $|E_z|^2$ reflectance for zigzag (light green; labelled "$R_z$") and armchair (dark green; labelled "$R_a$") terminations. Purple lines shows the reflectance difference (labelled "$R_a/R_z$") between the two terminations. (d) The *k*-space analysis on the out-coupling of TM polarization. The green circle represents the TM dispersion in the parallel plate waveguide. (e-f) The simulated refraction of TE and TM kink states through zigzag termination respectively. The right panel shows the experimentally captured field patterns. The white bars indicate the position of phase-arrayed dipoles. (g-h) The simulated refraction of TE and TM kink states through armchair termination respectively. The right panel shows the experimentally captured field patterns.